\documentclass[12pt]{article}
\usepackage{psfig}

\makeatletter
\def\vereq#1#2{\lower3pt\vbox{\baselineskip1.5pt \lineskip1.5pt
\ialign{$\m@th#1\hfill##\hfil$\crcr#2\crcr\sim\crcr}}}
\makeatother

\begin{document}

\begin{titlepage}
\begin{center}
\today     \hfill    LBNL-42246 \\
~{} \hfill UCB-PTH-98/45  \\
~{} \hfill hep-th/9810020\\

\vskip .1in

{\large \bf Seiberg Duality and $e^{+} e^{-}$ Experiments}%
\footnote{This work was supported in part by the U.S. 
Department of Energy under Contracts DE-AC03-76SF00098, in part by the 
National Science Foundation under grant PHY-95-14797.  HM was also 
supported by the Alfred P. Sloan Foundation and AdG by CNPq (Brazil).}

\vskip 0.3in

Andr\'e de Gouv\^ea,$^{1,2}$ Alexander Friedland,$^{1,2}$ and
Hitoshi Murayama$^{1,2,3}$

\vskip 0.05in

{$^1$\em Department of Physics, University of California\\
Berkeley, CA 94720}

\vskip 0.05in

{$^2$\em Theoretical Physics Group, Lawrence Berkeley National 
Laboratory\\ Berkeley, CA 94720}

\vskip 0.05in

{$^3$\em Theory Division, CERN\\
     Geneva, Switzerland}

\vskip 0.05in

\end{center}

\vskip .1in

\begin{abstract}
Seiberg duality in supersymmetric gauge theories is the claim that two 
different theories describe the same physics in the infrared limit.  
However, one cannot easily work out physical quantities in  
strongly coupled theories and hence it has been difficult to compare the 
physics of the electric and magnetic theories.  In order to gain more 
insight into the equivalence of two theories, we study the ``$e^{+} 
e^{-}$'' cross sections into ``hadrons'' for both theories in the 
superconformal window.  We describe a technique which allows us to
compute the cross 
sections exactly in the infrared limit.  They are indeed equal in 
the low-energy limit and the equality is guaranteed because of the
anomaly matching condition. 
The ultraviolet behavior of the total ``$e^{+} e^{-}$'' cross section 
is different for the two theories. We comment on proposed
non-supersymmetric dualities.  We also analyze the agreement
of the ``$\gamma\gamma$'' and ``$WW$'' scattering amplitudes in both
theories, and in particular try to understand if their equivalence can 
be explained by the anomaly matching condition.
    
\end{abstract}

\end{titlepage}

\newpage
\setcounter{footnote}{0}
\section{Introduction}

In the last few years, remarkable 
progress has been made on understanding supersymmetric gauge theories.  
An important part of the progress is the so-called Seiberg duality 
\cite{duality}, which states that two distinct supersymmetric gauge 
theories describe the same physics in the infrared 
limit.  In some cases both theories are in the non-Abelian Coulomb phase, 
with scale-invariant dynamics and no particle interpretation, {\it
  i.e.,}\/ asymptotic states cannot be defined.  In 
two-dimensions, there are techniques developed in conformal field
theories which allow the calculation of correlation functions of
conformal fields exactly, 
thanks to the infinite-dimensional extension of the conformal 
symmetry.  Unfortunately, much less is known about four-dimensional 
scale-invariant theories.

We attempt to extract some physical quantities from the proposed dual 
supersymmetric scale-invariant theories.  Since there is no particle 
interpretation for conformal fields, one cannot discuss $S$-matrices 
among quarks and gluons in these theories.  This situation is somewhat 
similar to the real-world QCD, where quarks and gluons cannot appear as 
asymptotic states, even though the reason for failure here is the color 
confinement rather than scale-invariant dynamics.  The similarity 
suggests the study of ``$e^{+} e^{-}$'' cross sections, which have 
direct physical meaning even in the absence of the precise knowledge of 
the excitations in the dynamics, and are known to be useful in studies 
of QCD. The problem then is to develop a technique to calculate 
these total cross sections exactly in scale-invariant supersymmetric theories.

We point out that the NSVZ beta-function allows us to determine the 
vacuum polarization amplitudes with global symmetry current operators 
exactly in the infrared limit for these theories, thanks to  
scale invariance.  By analytically continuing them to the time-like 
region, we obtain the exact ``$e^{+} e^{-}$'' cross sections even if the 
theory is strongly-coupled.  This allows us to compare results 
in the electric and magnetic theories and check the duality assumption. We 
find that the ``$e^{+} e^{-}$'' cross sections indeed agree between 
the proposed dual theories. We then show that the agreement between
the total cross sections is
guaranteed by the anomaly matching condition, which was required in
\cite{duality} to determine if the two theories are dual. 

On the other hand, checking that both theories yield the same ``$e^{+}
e^{-}$'' total cross sections is not enough to guarantee that the two
theories in questions are dual: {\it all} physical processes have to agree
in the infrared limit. We next investigate the
``$\gamma\gamma\rightarrow\gamma\gamma$'' and ``$WW\rightarrow WW$''
scattering amplitudes and, even though we
are unable to compute them exactly, address the issue of their
possible connection to the global anomalies.

\section{NSVZ $\beta$-function and Scale-invariant Theories}

In this section, we review the consistency checks on scale-invariant 
theories done originally in \cite{duality} and
\cite{magnetic_check}.  We do this review in order to make our
discussion of the two dual theories more complete. It also serves the 
purpose of defining some notation and deriving some results that will be 
used in the remainder of this letter.
We use the so-called NSVZ exact beta-function of 
the gauge coupling constant, which is given in terms of the one-loop 
beta function and the anomalous dimension of the matter fields.  We 
first determine the exact value of the anomalous dimensions in the
infrared  and then check 
that they agree with the expectation from superconformal symmetry, 
which relates the $U(1)_{R}$ charges to the conformal weights of the 
chiral superfields.  

The NSVZ exact beta-function for $N=1$ pure Yang--Mills theory 
was originally obtained heuristically by Jones 
\cite{Jones} and also 
by Novikov--Shifman--Vain\-shtein--Zakharov \cite{NSVZ1} using the 
instanton technique.  It was further generalized to theories with 
matter fields in \cite{NSVZ2}.  All of these results can now be 
obtained from the rescaling anomaly \cite{Nima2}.  The result in 
\cite{Nima2} is that this is the form of the beta-function one obtains 
in the Wilsonian effective actions for canonically normalized fields 
with manifestly supersymmetric and holomorphic regularizations.  It is 
even non-perturbatively exact in theories with an anomalous $U(1)$ 
symmetry but with no condensates \cite{Nima2,GM}.\footnote{This 
argument, however, assumes the existence of ultraviolet and infrared 
regulators which are invariant under the anomalous $U(1)$.} The NSVZ 
beta-function relates the running of the gauge coupling constant to 
the anomalous dimension factor $\gamma_{f}$ of the matter fields,
\begin{equation}
        \mu\frac{\rm{d}}{\rm{d}\mu}g^{2} = - g^{4}
        \frac{b_{0} + \sum_{f} T_{f} \gamma_{f}}{8\pi^{2} - C_{A} g^{2}},
        \label{eq:NSVZbeta}
\end{equation}
which can be analytically integrated to
\begin{equation}
        \frac{8\pi^{2}}{g^{2}(\mu)} + C_{A} \log g^{2}(\mu)
        = \frac{8\pi^{2}}{g^{2}(M)} + C_{A} \log g^{2}(M) + b_{0} \log 
        \frac{\mu}{M} - \sum_{f} T_{f} \log Z_{f}(\mu,M) ,
        \label{eq:NSVZ}
\end{equation}
where $\gamma_{f} = -\mu(d/d\mu) \log Z_{f}(\mu,M)$.  Here, 
$C_{A} \delta^{ab} = f^{acd} f^{bcd}$ is the (half of the) Dynkin 
index for the adjoint representation, 
$M$ is the ultraviolet cutoff, $\mu$ the renormalization scale, 
$b_{0} = 3 C_{A} - \sum_{f} T_{f}$ the one-loop beta-function 
coefficient, $T_{f} \delta^{ab} = \mbox{Tr}_{R_{f}} T^{a} T^{b}$ the 
(half of the) Dynkin index for the matter field $f$ in the 
representation $R_{f}$, and $Z_{f} (\mu, M)$ is the wave-function 
renormalization factor at the renormalization scale $\mu$ in the 
Wilsonian action.  For $SU(N_{c})$ gauge groups, $C_{A} = N_{c}$ and 
$T_{f}=1/2$ for the fundamental representation.

In the $N=1$ $SU(N_{c})$ QCD (SQCD), there are $N_{f}$ quarks $Q$ in the 
fundamental representation and $N_f$ anti-quarks $\tilde{Q}$ in the 
anti-fundamental representation. This will also be referred to as the
electric theory \cite{duality}.  The one-loop beta function 
coefficient is therefore $b_{0} = 3 N_{c} - N_{f}$, and $Z_{Q} = 
Z_{\tilde{Q}}$ independent of the flavor.  The claim in \cite{duality} 
is that the theory is scale-invariant if $\frac{3}{2}N_{c} < N_{f} < 3 
N_{c}$.  In order for the theory to be scale-invariant, the gauge 
coupling constant should not run.  Therefore the last two terms in 
Eq.~(\ref{eq:NSVZ}) must cancel. That is guaranteed if 
\begin{equation}
        Z_{Q} (\mu, M) = Z_{\tilde{Q}}(\mu, M) = \left( \frac{\mu}{M} 
        \right)^{(3N_{c}-N_{f})/N_{f}}.
        \label{eq:ZQ}
\end{equation}

For scale-invariance to be consistent with supersymmetry, the 
theory must be superconformal, which requires the conformal dimension 
of the chiral superfields to be given by $D = \frac{3}{2}R$, where $R$ 
is the (non-anomalous) $U(1)_{R}$ charge.  In SQCD, $R = (N_{f} - 
N_{c})/N_{f}$ for both $Q$ and $\tilde{Q}$.  Then the conformal 
dimension of the kinetic operator
\begin{equation}
        \int d^{4} x \, d^{4}\theta\, Q^{\dagger} e^{2V} Q
\end{equation}
is given by 
\begin{equation}
        (-4) + (+2) + 2 \frac{3}{2} \frac{N_{f}-N_{c}}{N_{f}}
        = \frac{N_{f} - 3 N_{c}}{N_{f}} .
\end{equation}
Therefore this operator must renormalize as 
$(\mu/M)^{(3N_{c}-N_{f})/N_{f}}$, which precisely coincides with the 
wave-function renormalization factor in Eq.~(\ref{eq:ZQ}).  In 
general, the exponent of $\mu/M$ is $-3 R + 2$.  The conformal 
dimension of the matter fields determined by the vanishing NSVZ 
beta-function is consistent with that determined by the superconformal 
symmetry from the $U(1)_{R}$ charge.\footnote{In theories with more 
global $U(1)$ symmetries, the scale invariance does not determine the 
$U(1)_{R}$ symmetry uniquely.  We do not know the anomalous 
dimensions of the matter fields uniquely either.} This consistency 
check was first performed by Seiberg \cite{duality}.

It is instructive to discuss the case where the infrared fixed point 
coupling constants are perturbative \cite{BZ}.  This occurs in the electric 
theory when $N_{f}$ is very close to but slightly below $3 N_{c}$, 
which is possible for large $N_{c}$.  Writing $N_{f}(1+\epsilon) = 3 
N_{c}$, we find
\begin{equation}
        Z_{Q} (\mu, M) = Z_{\tilde{Q}}(\mu, M) = \left( \frac{\mu}{M} 
        \right)^{\epsilon} = 1 + \epsilon \log \frac{\mu}{M} 
        + O\left(\epsilon\log \frac{\mu}{M}\right)^{2}.
\end{equation}
This expression can be compared to the perturbative 1-loop expression 
\begin{equation}
        Z_{Q} (\mu, M) = Z_{\tilde{Q}}(\mu, M) = 1+ \frac{g^{2}}{8\pi^{2}} 2 
        C_{2} \log \frac{\mu}{M} + O(g^{4}),
\end{equation}
where $C_{2} = (N_{c}^{2}-1)/2N_{c}$ is the second-order Casimir 
invariant for the (anti) fundamental representation.  The required 
anomalous dimension is obtained with a fixed-point coupling 
\cite{duality}
\begin{equation}
        \frac{g_{*}^{2}}{8\pi^{2}} = \epsilon \frac{N_{c}}{N_{c}^{2} - 1},
        \label{eq:fixed}
\end{equation}
up to $O(g^{4})$ corrections.  Note that $N_{c} g^{2}/8\pi^{2}$ is the 
relevant combination for large $N_{c}$ and $N_{c} g_{*}^{2}/8\pi^{2}$ 
is indeed small (order $\epsilon$) in the limit above.  This makes 
perturbation theory applicable in the entire energy range 
\cite{BZ,duality}, and justifies the omission of $O(g^{4})$ 
corrections a posteriori.  The coupling constant above is indeed the 
infrared fixed point value; the gauge coupling constant approaches the 
above value in the infrared limit from either larger or smaller 
ultraviolet couplings.  It also proves to be useful to consider limits 
$\epsilon \rightarrow +0$ ($N_{f} \rightarrow 3 N_{c} -0$) to check 
some of the results we will obtain later.  In this one-sided 
limit,\footnote{The limit $N_{f} \rightarrow 3 N_{c} + 0$
requires a different analysis because the theory in this case is free.  
For us, it is enough to discuss the one-sided limit.} the infrared fixed-point 
coupling can become arbitrarily small, and the lowest-order in 
perturbation theory should yield exact results.

We now proceed to check that the magnetic theory is also superconformal
\cite{magnetic_check}.  The 
magnetic theories \cite{duality} are given by $SU(\tilde{N}_{c})$ SQCD with 
$\tilde{N}_{c} = N_{f} - N_{c}$ and $N_{f}$ dual quarks $q$ and 
anti-quarks $\tilde{q}$ together with a meson field $M$, and the 
superpotential $W = 
\lambda M^{ij}\tilde{q}_{i} q_{j}$.\footnote{In this letter, we drop 
the dependence on the matching scale $\mu$ and assume that the meson field 
$M^{ij}$ has canonical dimension one.  This way, we can avoid 
confusion with the renormalization scale $\mu$, and the perturbative 
discussion on the meson Yukawa coupling $\lambda$ to the dual quarks 
is clearer.} The meson field $M^{ij}$ in the magnetic theory 
corresponds to the gauge-invariant composite operator $\tilde{Q}^{i} 
Q^{j}$ in the electric theory.  The magnetic theory is claimed to 
describe exactly the same physics as the electric theory in the 
infrared limit.  One implication of this is that the magnetic theory must also 
be scale-invariant in the infrared for $\frac{3}{2} N_{c} < N_{f} < 3 
N_{c}$.  The same consistency check done in the electric theory can 
be done for the $SU(\tilde{N_c})$ gauge coupling constant.  The wave-function 
renormalization factor of the dual quarks are determined to be, in the 
infrared,
\begin{equation}
        Z_{q} (\mu, M) = Z_{\tilde{q}} (\mu, M) = \left( \frac{\mu}{M} 
        \right)^{(3\tilde{N}_{c}-N_{f})/N_{f}}
        \label{eq:Zq}
\end{equation}
such that, according to the NSVZ formula Eq.~(\ref{eq:NSVZ}), the 
gauge coupling constant does not run.  This wave-function 
renormalization factor is consistent with the conformal dimension of 
the dual quarks expected from their $R$-charge and the superconformal 
symmetry. It is interesting to note that the form of the anomalous
dimension of $Q$ in the electric theory
and $q$ in the magnetic theory is the same, even though the magnetic
theory also includes meson fields. 

An additional check is that the Yukawa coupling of the meson field to 
the dual quarks should also not run. From the $R$-charge assignment 
of the meson field $R=2 (N_{f} - N_{c})/N_{f} = 2 \tilde{N}_{c} / 
N_{f}$ and superconformal symmetry, we need
\begin{equation}
        Z_{M} (\mu, M) = 
        \left( \frac{\mu}{M} \right)^{6 \tilde{N}_{c}/N_{f} - 2}
        \label{eq:ZM}
\end{equation}
exactly, in the infrared. One can easily check that 
$Z_{q} Z_{\tilde{q}} Z_{M} = 1$, which 
guarantees that the Yukawa coupling constant does not run:
\begin{equation}
        \lambda (\mu) = \lambda (M) Z_{q}^{-1/2} Z_{\tilde{q}}^{-1/2} 
        Z_{M}^{-1/2} = \lambda(M).
\end{equation}
This consistency check follows from the requirement that the 
total $U(1)_{R}$ charge of the superpotential is 2, as shown below.  
The general formula for the exponent of $\mu/M$ in the wave-function 
renormalization factor required from the superconformal symmetry is 
$-3 R + 2$, as discussed earlier.  The renormalization of the coupling 
is then given by $\mu/M$ to the power  
$-\frac{1}{2} \sum_{i=1}^{3} (-3 R_{i} + 2)$.  
The superpotential has $R$-charge 2, and hence $\sum_{i} R_{i} = 
2$.  This guarantees the absence of running for the Yukawa 
coupling.\footnote{For quartic terms in the superpotential, the 
coupling constants can be normalized by the ultraviolet cutoff to 
define dimensionless coupling constants, such as $\frac{\lambda}{M} 
\phi_{1} \phi_{2} \phi_{3} \phi_{4}$.  The analysis here shows that 
the coupling is renormalized by the ratio $\mu/M$ with the exponent 
$-\frac{1}{2} \sum_{i=1}^{4} (-3 R_{i} + 2) = -1$, and hence the 
operator becomes $\frac{\lambda}{\mu} \phi_{1} \phi_{2} \phi_{3} 
\phi_{4}$ with the new cutoff $\mu$.  Therefore, the {\it 
dimensionless}\/ coupling $\lambda$ also does not run.  
The same argument holds for higher dimensional operators 
in the superpotential.}

The infrared fixed point couplings are perturbative if $N_{f}$ is 
close to but slightly above $\frac{3}{2}N_{c}$; in other words, 
$N_{f} (1+\epsilon) = 3 \tilde{N}_{c}$.  Following the same discussion 
as before, the scale-invariance of the theory requires 
\begin{eqnarray}
        \lefteqn{Z_{q} (\mu, M) = Z_{\tilde{q}}(\mu, M)} \nonumber \\
        & & = \left( \frac{\mu}{M} 
        \right)^{\epsilon} = 1 + \epsilon \log \frac{\mu}{M} 
        + O\left(\epsilon\log \frac{\mu}{M}\right)^{2},
        \nonumber \\
        \lefteqn{Z_{M}(\mu,M) = \left( \frac{\mu}{M} 
        \right)^{-2 \epsilon} = 1 -2  \epsilon \log \frac{\mu}{M} + 
        O\left(\epsilon\log \frac{\mu}{M}\right)^{2}.}
\end{eqnarray}
Comparing the results above to perturbative 1-loop expressions
\begin{eqnarray}
        \lefteqn{Z_{q} (\mu, M) = Z_{\tilde{q}}(\mu, M)} \nonumber \\
        & & = 1+
        \left(\frac{g^{2}}{8\pi^{2}} 2 
        C_{2} - \frac{\lambda^{2}}{8\pi^{2}} N_{f} \right) \log 
        \frac{\mu}{M} + O(g^{4}, g^{2} \lambda^{2}, \lambda^{4}), \\
        \lefteqn{Z_{M} (\mu, M) = 
        1 - \frac{\lambda^{2}}{8\pi^{2}} \tilde{N}_{c} \log 
        \frac{\mu}{M} + O(g^{4}, g^{2} \lambda^{2}, \lambda^{4}),}
\end{eqnarray}
one finds the fixed point couplings:
\begin{eqnarray}
        \frac{g_{*}^{2}}{8\pi^{2}}& = & \epsilon 
        \frac{\tilde{N}_{c}}{\tilde{N}_{c}^{2}-1}
        \left(1 + 2 \frac{N_{f}}{\tilde{N}_{c}} \right) ,
          \\
        \frac{\lambda_{*}^{2}}{8\pi^{2}} & = & 2\epsilon 
        \frac{1}{\tilde{N}_{c}} ,
\end{eqnarray}
up to $O(g^{4}, g^{2} \lambda^{2}, \lambda^{4})$ corrections.  Again 
the omission of higher order terms is a posteriori justified when 
$\epsilon$ is small.

\begin{figure}[t]
\centerline{
  \psfig{file=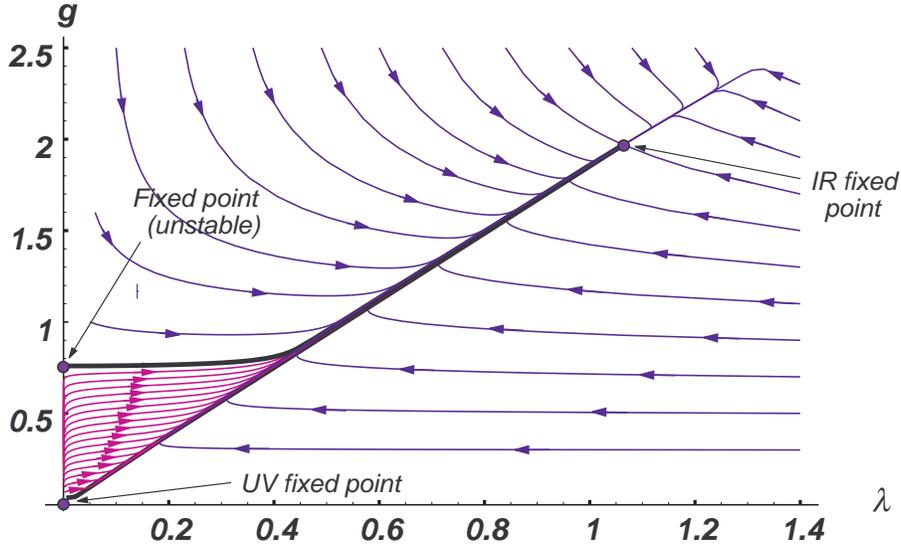,width=0.9\textwidth}
}
        \caption{The renormalization-group infrared flow of the
        coupling constants 
        $(\lambda,g)$  in the magnetic 
        SQCD, for $\tilde{N_{c}}= 7$, $N_{f} = 20$.  The infrared and
        ultraviolet  
        fixed points value are indicated.   
        All initial values shown in the plot flow to the infrared 
        fixed-point couplings.  The region bounded by the thick line is 
        asymptotically free; strictly speaking, the theory exists only within 
        this region.}
        \label{fig:flow}
\end{figure}

We depicted the renormalization-group flow of the coupling constants 
on the $(\lambda,g)$-plane for $\epsilon = 1/20$ in 
Fig.~\ref{fig:flow}.  For the region shown in the plot, where both
couplings are perturbative, all initial conditions flow to the fixed
point in the infrared.

One can also analyze the $(\lambda,g)$-plane to determine the region 
where the magnetic theory is asymptotically free in both coupling
constants. Keeping only order $\epsilon$ terms, the renormalization
group equation can be rewritten as 
\begin{eqnarray}
\frac{{\rm{d}}x}{\rm{d}\log\mu} & = & x(7x-2y), \\
\frac{{\rm{d}}y}{\rm{d}\log\mu} & = & -3y^2(\epsilon-y+3x),
\end{eqnarray}
where $x=\tilde{N_c}\lambda^2/8\pi^2$ and
$y=\tilde{N_c}g^2/8\pi^2$. Note that $x$ and $y$ are small
(order $\epsilon$), according to the expressions derived above for the 
fixed point couplings (the infrared fixed point is $x_{*}=2\epsilon$
and $y_{*}=7\epsilon$). 

There are 2 other fixed points:
$x_{0}=y_{0}=0$ and $x_{\epsilon}=0$,
$y_{\epsilon}=\epsilon$. $(x_{\epsilon},y_{\epsilon})$ is both
infrared and ultraviolet unstable, while 
$(x_0,y_0)$ serves as the ultraviolet fixed point (this was mentioned
briefly in \cite{magnetic_check}) if the couplings specified at some
intermediate scale $m$ belong to a particular region of the $(\lambda,g)$
plane. This region is indicated in Fig.~\ref{fig:flow}. To the best of 
our knowledge this analysis was not done elsewhere.

The form of asymptotically free region can be easily understood.
First note that there are two lines in the $(x,y)$-plane where d$x/$d$t$ and 
d$y/$d$t$ vanish, respectively: $2y=7x$ and $y=3x+\epsilon$. It is
clear that $(x_*,y_*)$ is the intersection of these two
lines. If $(x(m),y(m))$ is above the second line, $y$ ($g$) clearly flows to 
infinity in the ultraviolet, while if $(x(m),y(m))$ is
below the first line it is $x$ ($\lambda$) that blows up in the ultraviolet.

In the region between the two lines, both d$x/$d$t$ and d$y/$d$t$ are
negative, {\it i.e.,}\/ the infrared flow is towards {\it larger}
values of $(x,y)$. Note that points inside of this region are confined to it
when running towards the infrared. 

In light of the above discussion, determining the asymptotically free
region is equivalent to determining the set of points in the region
above that cannot be reached by outside points when flowing towards
the infrared. The boundaries of this region are determined
quantitatively, by looking at the infrared flow of points very close
to unstable fixed points but outside the region between the
two lines. It is clear from the infrared flow, for example, that 
$(x(m),y(m))=(x_{\epsilon}+\delta,y_{\epsilon}-\delta')$ (where
$\delta,\delta'$ are infinitesimal) cannot be reached by any
outside infrared flow. As a matter of fact, this point flows to
$(x_0,y_0)$ in the ultraviolet.

Note that numerically this behavior is somewhat ``masked,'' and it
looks as though all the infrared flow is almost horizontal 
until the $2y=7x$ line is reached. Then, it appears that the flow 
is along that line until the infrared fixed point is reached. The
reason for this is clear. Note that d$x/$d$t\propto x^2$
while d$y/$d$t\propto y^2\epsilon$. Since $\epsilon\ll 1$, the above
numerical behavior is obvious and the asymptotic free region is well
approximated by the triangle
$[(0,0),(0,\epsilon),(7/2\epsilon,\epsilon)]$ in the $(x,y)$-plane.
        
\section{Gauging Global Symmetries}

Scale-invariant theories do not admit particle interpretation for 
their conformal fields unless their conformal dimensions are those of
free fields.  Even though Seiberg duality 
states that the electric theory gives precisely the same physics as 
the magnetic theory in the infrared limit, it is still not clear to 
the authors what physical quantities can be compared between these 
superconformal theories.
The challenge is to identify suitable physical quantities of 
interest and to develop techniques to calculate them exactly in the 
infrared limit.  Unlike in two-dimensional space-time, techniques are 
not well developed to work out correlation functions of conformal 
fields in four dimensions (see, however, \cite{superconf}).

Historically, the study of the non-trivial gauge dynamics of QCD was done 
very effectively with $e^{+} e^{-}$ experiments.\footnote{The other type 
of useful experiment, deep inelastic scattering, cannot be discussed 
in the context of superconformal theories, because of the lack of 
bound-state hadrons and our 
inability to predict the parton distribution functions from first 
principles.  This problem exists even in QED if one takes the limit 
of massless electrons.} Here, a global symmetry ($U(1)_{EM}$) of  
QCD is gauged weakly and the gauge boson (photon) is produced 
off-shell from the $e^{+} e^{-}$ annihilation to create excitations in 
QCD. We would like to follow this program to gain more insight into 
the scale-invariant theories and especially their equivalence.

We first study the gauging of $U(1)_{B}$ in SQCD, and introduce 
``electrons'' which are charged under $U(1)_{B}$ but not under the 
SQCD gauge group.  We would then like to calculate the total cross 
section of creating excitations in SQCD from ``electron-positron'' 
annihilation.

At first sight, the cross sections do not appear to be the same in the 
electric and magnetic theories.  At the tree-level, one can easily 
compute the Drell ratio $R$ (cross sections normalized by the 
``point'' cross section $\sigma(e^{+} e^{-} \rightarrow \mu^{+} 
\mu^{-})$ where both ``electron'' and ``muon'' carry $U(1)_{B}$ charge 
unity).\footnote{Here and below, we employ a slightly modified definition of 
the ``point'' cross section which includes the production of both $\mu^{+} 
\mu^{-}$ and $\tilde{\mu}^{+} \tilde{\mu}^{-}$ for one chiral 
super-multiplet ({\it i.e.}\/, Weyl) rather than a full Dirac multiplet.  
The same comment applies to the $Q\bar{Q}$ production, which will include 
entire super-multiplets.  All the expressions are simpler with this 
definition.  If one wishes to go back to the traditional 
definition of the ``point'' cross section for a massless Dirac muon 
but no scalars, one should multiply our $R$ by a factor of 3/4.} One 
finds
\begin{equation}
        R \equiv \frac{\sigma(e^{+} e^{-} \rightarrow Q\bar{Q})}
        {\sigma(e^{+} e^{-} \rightarrow \mu^{+} \mu^{-})}
        = 2 N_{c} N_{f} \label{eq:UVRe}
\end{equation}
in the electric theory, while
\begin{equation}
        R \equiv \frac{\sigma(e^{+} e^{-} \rightarrow q\bar{q})}
        {\sigma(e^{+} e^{-} \rightarrow \mu^{+} \mu^{-})}
        = 2 \tilde{N}_{c} N_{f} \left(\frac{N_{c}}{\tilde{N}_{c}}\right)^{2}
        = 2 N_{c} N_{f} \frac{N_{c}}{N_{f}-N_{c}}
        \label{eq:UVRm}
\end{equation}
in the magnetic theory.  They clearly do not agree.  Of 
course, this is only a tree-level result and receives corrections at 
all orders in perturbation theory.  These results are only valid in 
the ultraviolet limit, where asymptotic freedom 
allows the approximation of $R$ by its tree-level value.  In the 
infrared limit, if the two theories describe the same physics, 
the Drell ratios must agree after 
corrections from all orders in perturbation theory are included.  The 
challenge is to find a way to calculate $R$ in both theories exactly 
in the infrared limit and compare them.  We describe such technique
below. A similar technique was
employed by Anselmi {\it et al} \/\cite{superconf} in the context of 
superconformal field theory.

The trick is to employ the NSVZ beta function for the running of the 
$U(1)_{B}$ coupling constant, and read off the vacuum polarization 
function from the beta function.  Since the NSVZ beta function sums 
up contributions from all orders in perturbation theory, the result on 
the vacuum polarization function also includes contributions from all 
orders.  One caveat is that it depends on the anomalous dimensions 
$\gamma_{f}$ which usually needs to be calculated in perturbation 
theory.  In the infrared limit, however, the anomalous 
dimension factors for SQCD are known exactly, as discussed in the previous
section, and one 
obtains the exact result for the vacuum polarization function.  Then it 
can be analytically continued to time-like momenta and its cut 
yields the cross section.

In general, the running of a $U(1)$ coupling is given by
\begin{equation}
        \frac{1}{e^{2}(\mu)} = \frac{1}{e^{2}(M)} + 
        \Pi(M^{2}) - \Pi (\mu^{2}),
        \label{eq:Pi}
\end{equation}
where the vacuum polarization amplitude\footnote{Our definition of the 
vacuum polarization amplitude $\Pi(Q^{2})$ does not include the 
$U(1)$ coupling constant.} $\Pi(Q^{2})$ depends on the Euclidean 
momentum $Q^{2} = - q^{2} = - g_{\mu\nu} q^{\mu}q^{\nu}$ with the 
metric $(+1, -1, -1, -1)$.  Comparing it to the NSVZ formula 
Eq.~(\ref{eq:NSVZ}), we find
\begin{equation}
        16\pi^{2}(\Pi(M^{2}) - \Pi(\mu^{2})) 
        = b_{0} \log \frac{\mu^{2}}{M^{2}} 
        - 2 \sum_{f} T_{f} \log Z_{f}(\mu,M) .
\end{equation}
If the functional dependence of $Z_{f}$ on $\mu$ is known, the 
analytic continuation of the vacuum polarization function gives the 
$e^{+} e^{-}$ cross section:
\begin{equation}
        R = 16\pi^{2} \frac{1}{2\pi i}(\Pi(-s+i\epsilon)-\Pi(-s-i\epsilon)) ,
\end{equation}
at the squared center-of-momentum energy $s$.  We apply this technique 
to the electric and magnetic theories in the infrared limit.

The first step is to calculate the running of the $U(1)_{B}$ gauge 
coupling constant.  We use the NSVZ formula with the infrared exact 
wave-function renormalization factors derived earlier 
({\it e.g.}\/ Eq.~(\ref{eq:ZQ})).  In 
the electric theory, we find\footnote{We omit the contribution of the 
``electromagnetic'' coupling to the wave-function renormalization 
factor and the contribution of the ``electrons'' to the beta-function.  
The former omission corresponds to the weak-coupling limit.  Both of 
these contributions can be easily incorporated if necessary; we ignore 
them for the simplicity of the discussion.  The weak-coupling limit, 
however, is a necessity not to spoil the superconformal invariance of 
the SQCD dynamics.}
\begin{eqnarray}
        \frac{8\pi^{2}}{e^{2}(\mu)} 
        &=& \frac{8\pi^{2}}{e^{2}(M)} - 2 N_{c} N_{f} \log 
        \frac{\mu}{M} - 2 N_{c} N_{f} \log \left( \frac{\mu}{M} 
        \right)^{(3N_{c}-N_{f})/N_{f}} \nonumber \\
        &=& \frac{8\pi^{2}}{e^{2}(M)} - 6 N_{c}^{2} \log 
        \frac{\mu}{M}.
\end{eqnarray}

The same calculation can be done in the magnetic theory.  The dual 
(anti) quarks carry baryon number $\pm N_{c}/\tilde{N}_{c}$, while 
the meson field has baryon number zero.  The application of the 
NSVZ formula gives
\begin{eqnarray}
        \lefteqn{
        \frac{8\pi^{2}}{e^{2}(\mu)} } \nonumber \\
        &=& \frac{8\pi^{2}}{e^{2}(M)} - 2 \tilde{N}_{c} N_{f} \left( 
        \frac{N_{c}}{\tilde{N}_{c}}\right)^{2} \log 
        \frac{\mu}{M} - 2 \tilde{N}_{c} N_{f} \left( 
        \frac{N_{c}}{\tilde{N}_{c}}\right)^{2} \log \left( \frac{\mu}{M} 
        \right)^{(3\tilde{N}_{c}-N_{f})/N_{f}} \nonumber \\
        &=& \frac{8\pi^{2}}{e^{2}(M)} - 6 N_{c}^{2} \log 
        \frac{\mu}{M}.
\end{eqnarray}
Therefore the running of the $U(1)_{B}$ coupling precisely agrees 
with what we obtained in the electric theory, despite quite 
distinct expressions at the intermediate stage of the calculations. 
This agreement makes the physical equivalence of the two theories much more 
direct and intuitive.  

Using the definition of the vacuum polarization function (\ref{eq:Pi}), one 
finds
\begin{equation}
        16\pi^{2}(\Pi(M^{2}) - \Pi(\mu^{2})) 
        = -6 N_{c}^{2} \log \frac{\mu^{2}}{M^{2}}.
\end{equation}
This is the {\it exact}\/ result in the infrared limit $\mu^{2} \ll 
\Lambda^{2}$ where $\Lambda$ is the dynamical scale of the gauge 
theory.  By analytically continuing $\Pi(\mu^{2})$ to the time-like 
region, the logarithm produces a cut, and one finds
\begin{equation}
        R = 6 N_{c}^{2}
        \label{eq:IRR}
\end{equation}
in the infrared limit ($s \ll \Lambda^{2}$).  Because the running of the 
$U(1)_{B}$ coupling is precisely the same in the infrared limit for the
electric and magnetic theories, this value of $R$ is also common for the
two theories.  This illustrates the equivalence of the two theories, at 
least in a limited fashion.  It is amusing to note that the result 
does not depend on $N_{f}$ at all.  From the point of view of  
electric theories, the result is the same for all values of $N_{f}$ 
from $\frac{3}{2}N_{c}$ to $3N_{c}$.  On the other hand from the point 
of view of the magnetic theories, the result depends only on $N_{c} = 
N_{f} - \tilde{N}_{c}$.  Another interesting point is that when the 
infrared fixed point coupling is perturbative, which happens when 
$N_{f}$ is close to but slightly less than $3N_{c}$ in the electric
theory, the exact result 
(\ref{eq:IRR}) and the tree-level result in the electric theory 
(\ref{eq:UVRe}) agree approximately.  They agree exactly in the 
limit $N_{f} \rightarrow 3 N_{c} - 0$ as noted earlier.  The same 
comment applies when $N_{f}$ is close to but slightly above 
$\frac{3}{2}N_{c}$; the exact result (\ref{eq:IRR}) and the tree-level 
result in the magnetic theory (\ref{eq:UVRm}) agree approximately.  
They agree exactly in the limit $N_{f} \rightarrow \frac{3}{2} 
N_{c} + 0$.

\begin{figure}[t]
\centerline{
\psfig{file=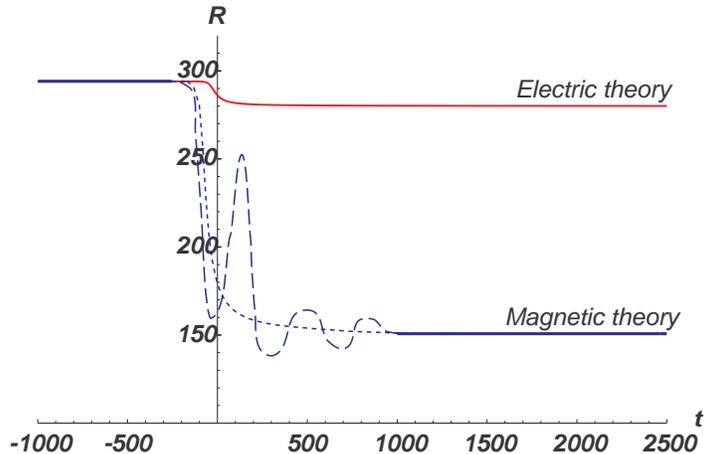,width=0.7\textwidth}
}
        \caption{The behavior of the Drell $R$-ratio for the electric 
        and magnetic SQCD as a function of
        $t=\log(\sqrt{s}/\Lambda)$, where $s$ is the 
        center-of-momentum energy squared and $\Lambda = M 
        \exp(-8\pi^{2}/g_{h}^{2}(M) b_{0})$. $M$ is the ultraviolet
        cut-off and $g_h$ is the holomorphic coupling. 
        We chose $N_{c}=7$ and $N_{f}=20$ so that the electric 
        theory is weakly-coupled for the entire $s$ range.  The magnetic 
        theory is strongly-coupled, and we cannot calculate $R$ for the 
        interpolating region $s \sim \Lambda^{2}$.  There are
        exact results for the limits $s\gg \Lambda^{2}$ and $s\ll 
        \Lambda^{2}$, however.  The dashed curves show two possible extreme 
        behaviors in the interpolating region.}
        \label{fig:R}
\end{figure}

\setcounter{footnote}{0}

For illustrative purpose, in Fig.~\ref{fig:R} the behavior of 
the $R$ is depicted as a function of $s$ in both the electric and 
magnetic theories 
for $N_{f}$ close to but slightly less than $3N_{c}$.  In this 
case the electric theory is weakly coupled and one can estimate $R$ 
throughout the whole energy range from $s \ll \Lambda^{2}$ to $s \gg 
\Lambda^{2}$.  We find:\footnote{This approximate formula assumes {\it 
local}\/ parton-hadron duality and that the gauge coupling constant in the 
time-like region $q^{2} =s$ is approximated by the corresponding one 
in the space-like region with $q^{2} = -s$.  This is a common 
assumption in high-energy $e^{+} e^{-}$ annihilations.  It is 
justified if the cross section is sufficiently smooth over a large range of 
$s$.  See, for instance, \cite{PS} for a pedagogical discussion of 
this point.}
\begin{equation}
        R (s) = 2 N_{f} N_{c} \left( 1 + \frac{g^{2}(s)}{8\pi^{2}} 2 
        C_{2} + O(g^{4})\right),
        \label{R}
\end{equation}
where $C_{2} = (N_{c}^{2} - 1)/2N_{c}$.  This expression smoothly 
interpolates the infrared and ultraviolet limits (Eqs.~(\ref{eq:IRR}) 
and (\ref{eq:UVRe})) given the fixed point coupling Eq.~(\ref{eq:fixed}).  
The magnetic theory is strongly coupled; however we calculated $R$ 
in two extreme limits: $s \ll \Lambda^{2}$ and $s \gg \Lambda^{2}$.  
We do not know how the two extreme values are interpolated.  Some 
possible behaviors are illustrated in the plot.  One sees a few 
remarkable facts in this plot.  First of all, the values of $R$ are the 
same in both theories in the low-energy limit, while they differ in 
the high-energy limit; this is precisely what is expected from the 
duality conjecture.  Second, the $R$-values ``drop'' at the dynamical 
scale (this was observed, in a different context, by the authors of
\cite{superconf}).  
The effect is especially prominent in the strongly coupled 
side (the magnetic theory for the particular example in the plot).  There 
is a plateau at the low-energy side, it goes through an interpolating 
region with either a smooth or wild behavior around the dynamical 
scale, and a new plateau sets in.  The new plateau is {\it lower}\/ 
than the other one.  The behavior which interpolates two plateaus is 
reminiscent of a threshold in QCD, {\it e.g.}\/, the $J/\psi$ region 
if it contains many bumps, or the $t\bar{t}$ threshold if it is smooth.  
However, thresholds in QCD always make the plateau {\it 
higher}\/.  The behavior in the plot appears, therefore, ``exotic'',
{\it i.\/e.}, contrary to common intuition.  
The drop is, of course, due to the decrease in the gauge 
coupling constant from the infrared fixed-point value to the 
asymptotically small coupling rather than the threshold effect of
exciting new degrees of freedom.

\setcounter{footnote}{0}

The same analysis holds for the $SU(N_{f})$ non-Abelian gauge groups.
The result in the electric theory is\footnote{Actually, one cannot 
gauge $SU(N_{f})$ in the SQCD because of the $(SU(N_{f})^{3})$ anomaly.  
One needs to add new particles to the theory to cancel the 
anomaly.  The comparison of the running coupling, however, is not 
affected by the presence of the additional fields thanks to the NSVZ 
formula.}
\begin{eqnarray}
        \lefteqn{
        \frac{8\pi^{2}}{e^{2}(\mu)} + N_{f} \log e^{2}(\mu) }
        \nonumber \\
        &=& \frac{8\pi^{2}}{e^{2}(M)} 
        + N_{f} \log e^{2}(M) 
        + (3N_{f}-\frac{1}{2}N_{c}) \log \frac{\mu}{M} 
        - \frac{1}{2}N_{c} \log 
                \left( \frac{\mu}{M} \right)^{\frac{(3N_{c}-N_{f})}{N_{f}}} 
        \nonumber \\
        &=& \frac{8\pi^{2}}{e^{2}(M)} + N_{f} \log e^{2}(M) 
        + \left(3N_{f} - \frac{1}{2} \frac{3 N_{c}^{2}}{N_{f}} \right) 
                \log \frac{\mu}{M}.
\end{eqnarray}
In the magnetic theory, the dual quarks as well as the meson fields 
contribute to the running.  We find
\begin{eqnarray}
        \lefteqn{
        \frac{8\pi^{2}}{e^{2}(\mu)} + N_{f} \log e^{2}(\mu) }
        \nonumber \\
        &=& \frac{8\pi^{2}}{e^{2}(M)} + N_{f} \log e^{2}(M) 
        + (3N_{f}-\frac{1}{2}\tilde{N}_{c}-\frac{1}{2}N_{f}) 
                \log \frac{\mu}{M} \nonumber \\
        & & 
        - \frac{1}{2}\tilde{N}_{c} 
                \log \left( \frac{\mu}{M} \right)^{(3\tilde{N}_{c}-N_{f})/N_{f}} 
        - \frac{1}{2} N_{f} 
                \log \left( \frac{\mu}{M} \right)^{-2(3\tilde{N}_{c}-N_{f})/N_{f}}
                \nonumber \\
        &=& \frac{8\pi^{2}}{e^{2}(M)} + N_{f} \log e^{2}(M) 
        + \left(3N_{f} - \frac{1}{2} \frac{3 N_{c}^{2}}{N_{f}} \right) 
                \log \frac{\mu}{M}.
\end{eqnarray}
Similarly to the $U(1)_{B}$ case, the running of the coupling 
constants precisely agree.  

One can introduce ``leptons'' coupled to the $SU(N_{f})$ gauge group, 
and discuss the ``lepton anti-lepton annihilation'' experiment.  The 
cross sections agree between the electric and magnetic theories in the 
limit $s\rightarrow 0$.  The second term in the bracket $\frac{1}{2} 
\frac{3 N_{c}^{2}}{N_{f}}$ gives the ``hadronic'' cross section from 
``lepton anti-lepton'' annihilation.  Note that the first term 
$(3N_{f})$ is related to the production of ``$W^{+} W^{-}$'' from the 
$s$-channel ``$W^{0}$-boson'' and is irrelevant for our discussion.  
We do not go into further details in this letter.

\section{Connection to Anomaly Matching}

We have seen in the previous section that the running of the gauge 
coupling constants for gauged global symmetries in SQCD precisely 
agree in the infrared limit between the electric and magnetic theories 
for the entire $\frac{3}{2}N_{c} < N_{f} < 3 N_{c}$ range, where the 
theories are scale-invariant.  This in turn guarantees that physical 
quantities, such as the $e^{+} e^{-}$ cross sections, are the same in
the two 
theories.  A natural question to address is whether the agreement of the cross 
sections poses a new constraint on duality or if it follows from
conditions imposed in \cite{duality}.  For the original arguments 
for duality to be sufficient, the agreement of the cross sections should 
not impose any new constraints but should follow from the
conditions already imposed.  We indeed find that the `t Hooft anomaly matching 
condition guarantees the agreement of the cross sections in these 
scale-invariant supersymmetric theories, and show why this is the case 
below. This was first pointed out, in 
a different context, in \cite{superconf}. 

As discussed in the previous sections, the wave-function 
renormalization factors in scale-invariant theories are given by 
$Z_{f}(\mu,M) = (\mu/M)^{-3R_{f}+2}$ because of superconformal 
symmetry.  First of all, it is interesting to check how the gauge 
coupling constant runs.  Using the NSVZ formula,
\begin{eqnarray}
        \lefteqn{
        \frac{8\pi^{2}}{g^{2}(\mu)} + C_{A} \log g^{2}(\mu) } \nonumber \\
        &=& \frac{8\pi^{2}}{g^{2}(M)} + C_{A} \log g^{2}(M) + (3 C_{A} - \sum_{f} 
        T_{f}) \log \frac{\mu}{M} - \sum_{f} T_{f} \log Z_{f}(\mu,M) 
        \nonumber \\
        &=& \frac{8\pi^{2}}{g^{2}(M)} + C_{A} \log g^{2}(M) 
        + 3\left(C_{A} + \sum_{f} (R_{f}-1) T_{f}\right) \log \frac{\mu}{M}.
\end{eqnarray}
The condition that the gauge coupling constant does not run is then 
given by $(+1)C_{A} + \sum_{f} (R_{f}-1)T_{F} = 0$.  This is the same 
condition as the requirement that $U(1)_{R}$ be anomaly free under the 
gauge group,\footnote{If the theory has additional global $U(1)$ 
summetries, any linear combination of $U(1)_{R}$ and the additional 
$U(1)$s still satisfy the same condition.  We cannot determine the 
$U(1)_{R}$ symmetry uniquely in this case.  But the equivalence of the 
NSVZ beta functions between electric and magnetic theories, which we 
will see below, still holds whatever $U(1)_{R}$ is chosen.} because 
the $U(1)_{R}$ charge of the gaugino is $+1$, and the fermionic 
component of the matter super-multiplet $f$ has $R$-charge $R_{f}-1$.

Now we apply the same analysis to the gauged global symmetry.  The 
NSVZ formula is precisely the same except that $g^{2}$ and the group 
theory factors $C_{A}$ and $T_{f}$ are now those of the gauged global 
symmetry.  The condition that the gauge coupling constant runs in the 
same way translates to the condition that the combination $\sum_{f} 
(R_{f}-1) T_{f}$ is the same between electric and magnetic theories, 
and this is nothing but the anomaly matching condition for $(U(1)_{R} 
G^{2})$, where $G$ is the global symmetry.  It is reassuring that the `t 
Hooft anomaly matching condition, checked in all proposed dualities, is enough 
to guarantee that physical observables, such as the cross sections
discussed in the previous section, are the same in the infrared
between the electric and magnetic theories in the 
superconformal window. 


It is noteworthy that the `t Hooft anomaly matching condition is 
presumably not enough to guarantee the equivalence of the cross 
sections in two scale-invariant theories without supersymmetry, such
as those  proposed in \cite{Terning}.  In the absence of supersymmetry, 
there is no reason for any one of the $U(1)$ symmetries of the theory 
to be related to the conformal transformation.  The combination $U(1)_{R} 
G^{2}$ was important in the supersymmetric theories because $U(1)_{R}$ 
is in the same multiplet as the dilation and hence is indeed related 
to the dilation anomaly ({\it i.e.}\/, running coupling constant) of 
the $G$ coupling.  In non-supersymmetric theories, it is quite 
possible that the check of `t Hooft anomaly matching is far from 
enough to guarantee that the two theories describe the same infrared
physics. The agreement 
of the cross sections considered here should be checked as an 
independent requirement for duality.

\section{Other Physical Observables}

We have seen that one can gauge global symmetries in SQCD and 
compare the running of the couplings in the electric and magnetic 
theories.  We could furthermore calculate the ``$e^{+} e^{-}$'' cross
sections in the infrared 
exactly in both theories and check that they are indeed the same.  Even though 
these examples made the physical equivalence of two theories more 
manifest and explicit, they are only a few out of an infinite number of 
physical observables.  All of them must be the same 
in the infrared when comparing the electric and magnetic theories if these
are indeed dual.
It is also important to see if the agreement of all observables 
follows from the `t Hooft anomaly matching condition or other consistency 
checks already done in the literature. The problem is that there
are only very limited known methods of calculating other physical observables 
in theories with scale-invariant dynamics in four dimensions, unlike in
the two dimensional case. 

For instance, one can try to calculate the ``light-by-light 
scattering'' cross sections of the $U(1)_{B}$ ``photons,'' induced by 
loops of quarks.  The effective operator
\begin{equation}
        C \int d^{4} \theta W_{\alpha} W^{\alpha} \bar{W}_{\dot{\alpha}} 
        \bar{W}^{\dot{\alpha}}
\end{equation}
can be defined with a suitable infrared cutoff $\mu_{IR}$.  At the lowest 
order in perturbation theory, the coefficient $C$ is proportional to $2 
N_{f} N_{c}$ in the electric theory and $2 N_{f} \tilde{N}_{c} 
(N_{c}/\tilde{N}_{c})^{4}$ in the magnetic theory; they 
are clearly different.  There is, as of today, no known powerful
technique which would allow one to work 
out such a coefficient {\it exactly}\/ in superconformal
theories.

In this section, we will develop a very simple argument that allows
one to constrain the form of $C$ and other related physical quantities.
The hope is that these coefficients will be calculated
exactly using the machinery of superconformal field theories, similar
to what was accomplished in \cite{superconf}. By constraining the form 
of the exact answer, it might be possible to gain some insight towards 
performing these computations,  and, hopefully, to decide if the anomaly
matching condition is enough to explain the agreement of these
physical observables. 

The argument proceeds as follows: we have already mentioned that 
lowest-order perturbation theory agrees with the exact answer in the
limit $N_f\rightarrow3N_c-0$ for the electric theory and
$N_f\rightarrow3/2N_c+0$ for the magnetic theory. We will use these
limits to constrain the exact result.

Returning to the total ``$e^{+}e^{-}$'' cross section,
the exact answer must be $2N_cN_f=6N_{c}^{2}$ for
$N_f\rightarrow3N_c-0$ and $2\tilde{N_c}N_f(N_c/\tilde{N_c})^2=6N_{c}^{2}$
for $N_f\rightarrow3/2N_c+0$. Further requiring that the exact 
answer be written in terms of the global anomalies of the theory
(which are guaranteed to be the same by anomaly matching), an obvious
candidate appears: $-3(U(1)_{R}U(1)_{B}^{2})=6N_{c}^{2}$. 

This candidate also satisfies another important constraint: the 
answer has to be invariant under simultaneous rescaling of the $U(1)_B$
gauge coupling by $n$ and the charge assigned to individual fields by
$1/n$, that is if $e\rightarrow ne$ and $q\rightarrow q/n$ physical
results should remain the same. Note that $\sigma(\gamma\rightarrow
hadrons)\propto e^2$ and the final answer must therefore include
$U(1)_B$ charges squared in order to guarantee this invariance.

By this sort of argumentation one cannot say that the correct
answer has been determined, but an acceptable candidate has certainly
been found. The exact
answer, which was derived from superconformal invariance and the NSVZ
exact beta function, in fact agrees with the naive guess above.

We now turn to the amplitude of
$\gamma\gamma\rightarrow\gamma\gamma$, and start from the same set of
assumptions: ({\it i}) the theories yield the same coefficient in the
infrared; ({\it ii}) the exact result must agree with the lowest order
perturbation theory in the two $N_f$ limits above; ({\it iii}) these
coefficients can be written in terms of the global anomalies of the
theory; ({\it iv}) the coefficient must be proportional to the $U(1)_B$ charge 
to the fourth power. All global anomalies can be easily computed and
they are \cite{duality}
\begin{eqnarray}
(SU(N_f)^{3}) & = & N_{c}d^{(3)}(N_f), \\
(SU(N_f)^{2}U(1)_{R}) & = &-{N_c^2\over{N_f}}d^{(2)}(N_f), \\
(SU(N_f)^{2}U(1)_{B}) & = & N_cd^{(2)}(N_f), \\
(U(1)_R) & = & -N_c^2-1, \\
(U(1)_R^3) & = & N_c^2-1-2{N_c^4\over{N_f^2}},\\
(U(1)_B^{2}U(1)_{R}) & = & -2N_c^2,
\end{eqnarray}
where $d^{(2)}(N_f)=1/2 \delta^{ab}$ and
$d^{(3)}(N_f)={\rm{Tr}}(T^a\{T^b,T^c\})=d^{abc}$ 
($a,b,c$ are $SU(N_f)$ indices, $T^a$ are $SU(N_f)$ generators).  

According to the assumptions above, a candidate for the exact result
must be:
\begin{eqnarray}
C & \propto & 
(U(1)_B^{2}U(1)_{R})^2 f + 
\nonumber \\
 &  & (SU(N_f)^{2}U(1)_{B})^4 g
+  \nonumber \\
&  & (SU(N_f)^{2}U(1)_{B})^2(U(1)_B^{2}U(1)_{R}) h, 
\end{eqnarray}
where $f,g,h$ are arbitrary functions of
\begin{displaymath}
(SU(N_f)^{3}), 
(SU(N_f)^{2}U(1)_{R}), 
(U(1)_R),(U(1)_R^3)
\end{displaymath}
and
\begin{displaymath}
(SU(N_f)^{2}U(1)_{B})^2/(U(1)_B^{2}U(1)_{R}).
\end{displaymath}
These will be referred to as the ``invariant anomalies'', because they 
are not sensitive to rescalings of the $U(1)_B$ charge assignments. 
Note that, in order
to make sense out of the $SU(N_f)$ anomalies, one has to sum over all
$SU(N_f)$ indices. 

Finally, conditions on $f$, $g$ and $h$ have to imposed such that 
\begin{displaymath}
C(N_f=3N_c) \propto 6N_c^2
\end{displaymath}
and
\begin{equation}
\label{conds}
C(N_f=3/2N_c) \propto 24N_c^2,
\end{equation}
for any value of $N_c$, according to assumption ({\it ii}). There is no
linear combination of integer products of the invariant anomalies that 
can meet these conditions. There are, of course, less ``obvious''
candidates. 

Note that exact infrared results, whatever they might be, are
functions exclusively of $N_c$ and $N_f$ (there are no other
independent parameters in the theories), and it is always possible to
write $N_c$ and $N_f$ in terms of a group of anomalies. One can, 
for example, take the $(U(1)_R)$ and the
$(SU(N_f)^{2}U(1)_{B})^2/(U(1)_B^{2}U(1)_{R})$ anomalies and simply solve for
$N_c$ and $N_f$:
\begin{eqnarray}
\label{NCNF}
N_c &=& \sqrt{-(U(1)_R)-1}, \nonumber \\
N_f &=& \sqrt{1-8(SU(N_f)^{2}U(1)_{B})^2/(U(1)_B^{2}U(1)_{R})}. 
\end{eqnarray}  
Having done that, it is easy to come up with a (simple) function of $N_f$ and
$N_c$ that satisfies the conditions Eqs.~(\ref{conds}). $54 N_c^4/N_f^2$
clearly does the job, and a candidate for the exact answer that
satisfies the conditions imposed above is 
\begin{equation}
C\propto \frac{27}{2}
{(U(1)_B^{2}U(1)_{R})^2\over{1-8(SU(N_f)^{2}U(1)_{B})^2/(U(1)_B^{2}U(1)_{R})}},
\end{equation}
where we used Eq.~(\ref{NCNF}). 

Next we address, within the same spirit, the
``$W_LW_L\rightarrow W_LW_L$'' scattering amplitude generated via
quark loops. The subscript $L$ refers to the $SU(N_f)$ group under
which $Q$ and $q$ transform ($\tilde{Q}$ and $\tilde{q}$ transform
under $SU(N_f)_R$).
Note that the coefficient of the amplitude is proportional to
${\rm{Tr}}(\{T^a,T^b\}\{T^c,T^d\})$ ($T^a$ are $SU(N_f)$ generators), 
which can be written in terms of the $SU(N_f)$ tensors present in 
the anomalies, namely $\delta^{ab}$ and $d^{abc}$. A candidate for the 
exact coefficient is 
\begin{equation}
C_{W_LW_LW_LW_L}\propto
{\rm{Tr}}(\{T^a,T^b\}\{T^c,T^d\})\frac{N_c^2}{N_f}=
\left(d^{abe}d^{cde}+\frac{1}{2N_f}\delta^{ab}\delta^{cd}\right )
\frac{N_c^2}{N_f}.
\end{equation}
The expression above can easily be written in terms of the anomalies
with the help of Eqs.~(\ref{NCNF}) and the fact that 
\begin{eqnarray}
\frac{\delta^{ab}}{2N_f} &=& \frac{-(SU(N_f)^2U(1)_R)^{ab}}{N_c^2} 
= \frac{-(SU(N_f)^2U(1)_R)^{ab}}{-(U(1)_R)-1},
\nonumber \\
d^{abc} &=& \frac{(SU(N_f)^3)^{abc}}{N_c} = 
\frac{(SU(N_f)^3)^{abc}}{\sqrt{-(U(1)_R)-1}}.
\end{eqnarray}
Note that the expression for $C_{W_LW_LW_LW_L}$ in terms of anomalies
only is highly non-trivial and involves, for example, inverse square-roots of
polynomials of anomalies.

As one last example, we mention $W_LW_L\rightarrow W_RW_R$
scattering. The amplitude for this process must vanish at $N_f=3N_c-0$ 
and be proportional to $N_f$ (remember that for the magnetic theory
there are the meson fields that transform under both $SU(N_f)$ flavor
groups) at $N_f=3/2N_c+0$. Therefore the candidate for the exact
result cannot be a ratio of $N_c$ and $N_f$ to arbitrary powers (as in 
the previous cases), but has to be, for example, proportional to
$3N_c-N_f$. We do not write the candidate in terms of
anomalies, but it is clearly possible.

Note that, since it was shown that any function of $N_c$ and $N_f$ can 
be reexpressed in terms of anomalies (as, for example, in
Eqs.~(\ref{NCNF})), it is 
always possible to write exact infrared results in terms of
anomalies. It is, however, unclear whether the agreement between the
electric and magnetic theories is guaranteed by the anomaly matching
condition. Given the expressions outlined above, it is at best
nonintuitive that  future superconformal field theory exact
computations  will yield highly
non-trivial functions of the anomalies as their results, especially
because  
the candidates are very simple functions of $N_c$ and $N_f$. 
The situation outlined above should be
contrasted to the ``$e^+e^-$'' total cross sections, described in 
the previous two sections.

Finally, it is clear that the ``candidates'' pointed out above are
by no means guaranteed to be the exact answer. They
agree with the exact answer in the limit when either the electric or
the magnetic theory is arbitrarily weak ($N_f=3N_c-0, N_f=3/2N_c+0$), 
but other functions of $N_c$ and $N_f$ also have the same
property. More constraints can be imposed on our candidates
if one tries to analyze the superconformal theories when they
are perturbative throughout the entire energy range
($N_f(1+\epsilon)=3N_c$ for the electric theory). In this case, next
order in perturbation theory can be used and, because $g^2$ and
$\epsilon$ are related in the infrared (see {\it e.g.}\/
Eq.(\ref{eq:fixed})), one can compare the order $g^2$
1-loop correction to the order $\epsilon$ correction obtained from the 
candidate exact result. Note that this is indeed the case for the perturbative
expression for $R$, Eq.~(\ref{R}).

Such a highly non-trivial check would be a very strong indication that one has
indeed found the exact infrared answer for a given observable. We do
not perform the next-order (2-loop) calculation for the amplitudes discussed
above; it goes beyond the scope of our letter.  

In summary, we hope we have raised and addressed in a limited fashion 
the following question: are all (or some)
physical observables in superconformal field theories (in four
dimensions) related to anomalies? If this indeed turns out to be the
case, anomaly matching would be enough to guarantee the physical
equivalence of two superconformal theories, which would shed light 
towards a ``physical'' understanding of Seiberg's dualtity
statements. On the other hand, if it is proven that there are physical 
observables which are not related to anomalies, then their exact
computation would prove to be an extra duality condition.

\section{Conclusion}

In this letter we tried to verify that Seiberg's duality conjecture is
indeed correct. Such a check is at best non-trivial, given that one
would have to perform computations in (very) strongly coupled
theories. 

To bypass this problem we have concentrated on the
non-Abelian Coulomb phase, where both theories are superconformal and
there is hope that some exact results can be obtained. The first
challenge in this phase is to come up with appropriate physical
observables that can be readily compared in both the electric and
magnetic theories, especially because conformal theories do not, in
general, allow any particle interpretation. 

We have described total cross sections, related to the process of
gauging the global symmetries of the theories, that can be used to check the
duality conjecture. Some exact results were obtained and they agree
between the magnetic and the electric theories.

Next we checked if the equality of these total cross
sections poses any new duality conditions. For SQCD the anomaly
matching condition is enough to guarantee that the total cross 
sections mentioned above are the same, thanks to
superconformal invariance. We pointed out that the situation is different
in non-supersymmetric dualities, and that the equalities of such
cross sections should be included as an extra constraint.

Finally we analyzed other physical observables and tried to determine
if their agreement between the electric and magnetic theories could
also be explained by the anomaly matching condition. We found that
that the functional form of the candidate exact result is highly
non-trivial when expressed in terms of the global anomalies of the
theories. It is not clear that superconformal field theory
computations would yield such results.

It is, therefore, possible that the exact results do not have any
relation to the global
anomalies of the theory, and the fact that all physical processes
agree for both theories is not guaranteed by Seiberg's criteria
alone.

More work is clearly required to fully understand this issue.

\section*{Acknowledgements} 
We thank John Terning for useful discussions and comments on the
draft.  HM thanks Nima Arkani-Hamed, Csaba Csaki, Ben Grinstein,
Markus Luty, John March-Russell, and Lisa Randall. HM also thanks the
hospitality of Theory Group at CERN, where some of the research was
conducted.  This work was supported in part by the U.S. Department of
Energy under Contracts DE-AC03-76SF00098, in part by the National
Science Foundation under grant PHY-95-14797.  HM was also supported by
Alfred P. Sloan Foundation, and AdG by CNPq (Brazil).


\begin{thebibliography}{99}

        \bibitem{duality} N. Seiberg, {\sl Nucl. Phys.}\/ {\bf B435}, 129 
        (1995), hep-th/9411149.

        \bibitem{magnetic_check} I. Kogan, M. Shifman and
          A. Vainshtein, {\sl Phys. Rev.}\/ {\bf D53}, 4526 (1996), hep-th/9507170.

        \bibitem{Jones} D.R.T. Jones, {\sl Phys. Lett.}\/ {\bf 123B}, 45 (1983).
        
        \bibitem{NSVZ1} V.A. Novikov, M.A. Shifman, A.I. Vainshtein, and V.I. 
        Zakharov, {\sl Nucl. Phys.}\/ {\bf B229}, 381 (1983).
        
        \bibitem{NSVZ2} V.A. Novikov, M.A. Shifman, A.I. Vainshtein, V.I. 
        Zakharov, {\sl Nucl. Phys.}\/ {\bf B260}, 157 (1985).
        
        \bibitem{Nima2} N. Arkani-Hamed and H. Murayama, hep-th/9707133. 
        
        \bibitem{GM} M. Graesser and B. Morariu, {\sl Phys. Lett.}\/ {\bf 
        B429}, 313 (1998), hep-th/9711054.
        
        \bibitem{BZ} T. Banks and A. Zaks, {\sl Nucl. Phys.}\/ {\bf B196}, 
        189 (1982). 

        \bibitem{superconf} D. Alselmi {\it et al}, 
        {\sl Nucl. Phys.}\/ {\bf B526}, 543 (1998), hep-th/9708042.
        
        \bibitem{PS} M. E. Peskin and D. V. Schroeder, ``An Introduction to 
        Quantum Field Theory,'' Reading, Addison-Wesley (1995),
        Chapter 18.
        
        \bibitem{Terning} J. Terning, {\sl Phys. Rev. Lett.}\/ {\bf 80}, 
        2517 (1998), hep-th/9706074.
\end{thebibliography}
\end{document}